\documentclass[journal]{IEEEtran}
\IEEEoverridecommandlockouts
\usepackage{cite}
\usepackage{amsmath,amssymb,amsfonts}
\usepackage{algorithmic}
\usepackage{graphicx}
\usepackage{textcomp}
\usepackage{xcolor}
\usepackage{multirow}
\usepackage{array}
\usepackage{color}
\usepackage{verbatim}
\usepackage{CJKutf8}
\begin{document}

\title{Belief Information based Deep Channel Estimation for Massive MIMO Systems\\
}

\author{Jialong Xu,~\IEEEmembership{Member,~IEEE,}
        Liu Liu,~\IEEEmembership{Member,~IEEE,}
        Xin Wang,~\IEEEmembership{Member,~IEEE,}
        and~Lan Chen,~\IEEEmembership{Senior Member,~IEEE}
\thanks{\textit{(corresponding author: Jialong Xu)}}
\thanks{Jialong Xu, Liu Liu, Xin Wang, and Lan Chen are with DOCOMO Beijing Communications
Laboratories, Co. Ltd., No. 2 Kexueyuan South Road, Beijing, 100191, China. (e-mails: \{xujl, liul, wangx, chen\}@docomolabs-beijing.com.cn).}
}


\maketitle

\begin{abstract}
In the next generation wireless communication system, transmission rates should continue to rise to support emerging scenarios, e.g., the immersive communications. From the perspective of communication system evolution, multiple-input multiple-output (MIMO) technology remains pivotal for enhancing transmission rates. However, current MIMO systems rely on inserting pilot signals to achieve accurate channel estimation. As the increase of transmit stream, the pilots consume a significant portion of transmission resources, severely reducing the spectral efficiency. In this correspondence, we propose a belief information based mechanism. By introducing a plug-and-play belief information module, existing single-antenna channel estimation networks could be seamlessly adapted to multi-antenna channel estimation and fully exploit the spatial correlation among multiple antennas. 
Experimental results demonstrate that the proposed method can either improve $1 \sim 2$ dB channel estimation performance or reduce $1/3 \sim 1/2$ pilot overhead, particularly in bad channel conditions.  
\end{abstract}


\section{Introduction}

Compared to the existing fifth-generation (5G) wireless communication system, 6G will be designed to support significantly higher throughput, accommodating potential high data-rate applications, e.g., the immersive extended reality (XR), holographic communications, and the digital twin. Building on the success the orthogonal frequency division multiplexing (OFDM) and multiple-input multiple-output (MIMO) employed in 4G and 5G, these technologies may be further enhanced to meet the increased throughput demand of 6G. 

Pilot-assisted channel estimation is designed to mitigate wireless channel impairment using pilot signals in existing wireless communication systems. However, as the number of required required data streams increases, the proportion of pilot signals in transmission resources  also rises, reducing spectral efficiency (SE) and neutralizing the benefit of larger MIMO configurations.

In recent years, the success of deep learning (DL) in the computer vision domain and natural language processing domain enlightened researchers to redesign DL-based algorithms to enhance conventional communication modules, e.g., channel state information (CSI) feedback \cite{xu2022deep} and beam management \cite{xu2023performance}, or to design the entire communication system with an end-to-end manner, e.g., joint source-channel coding \cite{xu2021wireless} and joint transceiver designs\cite{ye2021deep}. Meanwhile, benefiting from the similarity between the channel estimation task, which maps the channel estimated at pilot positions to the full estimated channel, and the super-resolution task, which maps a low-resolution image to a high-resolution image, DL is introduced to channel estimation \cite{soltani2019deep}. Then efficient neural network architectures are introduced and redesigned for DL-based channel estimation with the consideration of wireless channel characteristics. For instance, the residual network, the transformer, and the multi-layer perceptron (MLP)-mixer inspire the design of ReEsNet\cite{li2019deep}, Channelformer\cite{luan2023channelformer} and CMixer \cite{chen2024channel}, respectively, to improve the channel estimation accuracy.

However, the aforementioned DL-based channel estimation methods typically assume the channel estimation for the single antenna, thereby overlooking the spatial relationship among multiple receiving antennas. In this correspondence, we propose a belief information based mechanism for multiple-antenna channel estimation. Inspired by the maximum ratio combination (MRC) algorithm, which assigns greater weight to stronger signals and less to weaker ones, this mechanism uses multi-antenna belief information to assign different weights to various spatial features within the channel estimation network. This approach fully leverages the correlation among multiple antennas, enhancing overall channel estimation accuracy."

The rest of the correspondence is organized as follows. Section II introduces the system model for channel estimation of multiple receiving antennas. A belief information based mechanism is proposed in Section III. Section IV evaluates the proposed method in terms of channel estimation accuracy, pilot overhead reduction, storage overhead and computational complexity. Finally, Section V concludes our work.
 
\section{System Model}
\begin{figure*}[t]
\centering
\includegraphics[width=2\columnwidth]{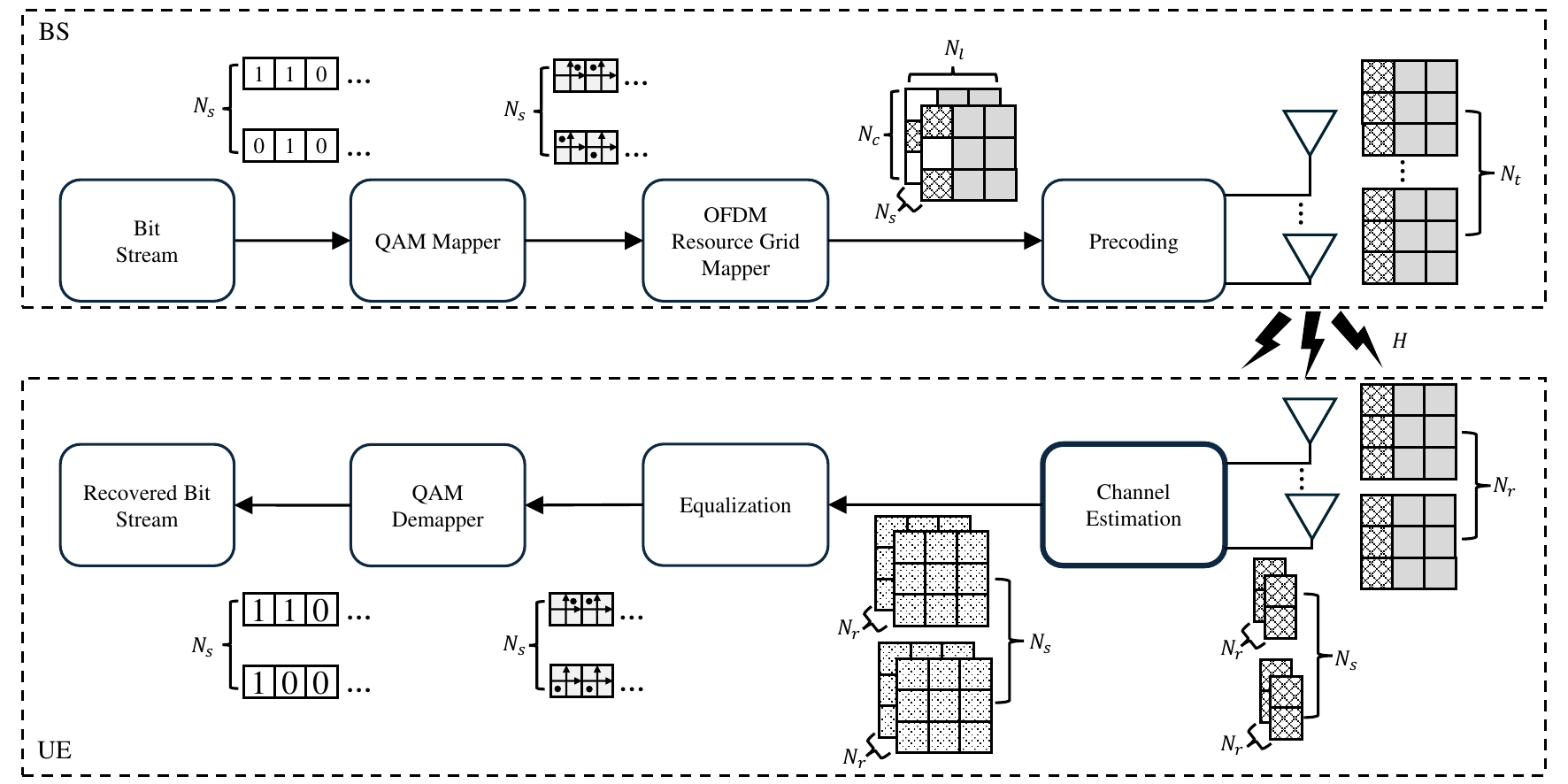}
\caption{The downlink transmission of the massive MIMO-OFDM communication system.}
\label{Fig:sys_model}
\end{figure*}
As shown in Fig.~\ref{Fig:sys_model}, we consider the downlink transmission of the massive MIMO-OFDM system across $N_{c}$ subcarriers. The base station (BS) and the user equipment (UE) are equipped with $N_{t}$ antennas and $N_{r}$ antennas, respectively. Each OFDM slot consists of $N_{l}$ OFDM symbols. For the $i$-th subcarrier, the downlink channel is represented as $H^{(i)} \in \mathbb{C}^{N_r \times N_t}$. 

At the BS, the transmitted bit streams $B \in \{0, 1\}^{N_{s} \times R \times M}$ is mapped to the data stream signals $X_{D} \in \mathbb{C}^{N_{s} \times R}$ by the quadrature amplitude modulation (QAM) of order $2^M$, where $R$ represents the available OFDM resources for each data stream and $N_{s}$ represents the number of data steams. Then the pilot signals $X_{P} \in \mathbb{C}^{N_{s} \times N_{cp} \times N_{lp}}$ are introduced and combined with the data stream signals $D$ to form the mapping signals $X \in \mathbb{C}^{N_{s} \times N_{c} \times N_{l}}$ by the OFDM resource grid mapper. In the precoding stage, The transmitted signal at BS antennas is expressed as:
\begin{equation}
S = PX \in \mathbb{C}^{N_{t} \times N_{c} \times N_{l}},
\end{equation}
where the precoding matrix is $P \in \mathbb{C}^{N_{t} \times N_{s}}$, which could be calculated according to the CSI feedback in frequency-division duplex (FDD) mode or the channel reciprocity in time-division duplex (TDD) mode. To concentrate on the channel estimation task, the accurate CSI is assumed for the precoding process. 

The transmitted signal is corrupted by the wireless channel and the additive white Gaussian noise (AWGN), which can be expressed as:
\begin{equation}
Y = HS + N = HPX + Z \in \mathbb{C}^{N_{r} \times N_{c} \times N_{l}},
\end{equation}
where the OFDM-MIMO channel and the AWGN noise at the receiver are represented as $H \in \mathbb{C}^{N_{r} \times N_{t} \times N_{c} \times N_{l}}$ and $Z \in \mathbb{C}^{{N_{r} \times N_{c} \times N_{l}}}$, respectively. The equivalent channel coefficient at the UE to be estimated becomes:
\begin{equation}
G = HP \in \mathbb{C}^{N_{r} \times N_{s} \times N_{c} \times N_{l}}.
\end{equation}

It is worth noting that the pilot signals are inserted into the OFDM resources with known subcarrier indices and slot indices. Hence the received signal at pilot positions is expressed as:
\begin{equation}
Y_{p} = G_{p}X_{p}+N_{p} \in \mathbb{C}^{N_{r} \times N_{cp} \times N_{lp}}.
\end{equation}
where $G_{p} \in \mathbb{C}^{N_{r} \times N_{s} \times N_{cp}\times N_{lp}}$ and $N_{p} \in \mathbb{C}^{N_{r} \times N_{cp}\times N_{lp}}$. Based on the known $Y_{p}$ and $X_{p}$, the equivalent channel $\hat{G}_{p} \in $ estimated at pilot positions is expressed as:
\begin{equation}
\hat{G}_{p} = f_{p}(Y_{p}, X_{p}),
\end{equation}
where $f_{p}: \mathbb{C}^{N_{r} \times N_{cp} \times N_{lp}} \times \mathbb{C}^{N_{s} \times N_{cp} \times N_{lp}} \rightarrow \mathbb{C}^{N_{r} \times N_{s}\times N_{cp} \times N_{lp}}$ represents the channel estimation function at pilot positions. Then the equivalent channel $\hat{G}$ can be inferred from the partial $\hat{G}_{p}$:
\begin{equation}
\hat{G} = f_{e}(\hat{G}_{p}),
\end{equation}
where $f_{e}: \mathbb{C}^{N_{r} \times N_{s} \times N_{cp} \times N_{lp}} \rightarrow \mathbb{C}^{N_{r} \times N_{s} \times N_{c} \times N_{l}}$ represents the expansion or the interpolation function. The whole channel estimation consists of the estimation function $f_{p}(\cdot)$ and the expansion function $f_{e}(\cdot)$. Traditional channel estimation methods often adopt the least square (LS) as the $f_{p}(\cdot)$, and linear (LIN) interpolation, nearest-neighbor (NN) interpolation, and linear minimum mean square error (LMMSE) interpolation as the $f_{e}(\cdot)$.

After the channel estimation, the estimated channel $\hat{G}$ is employed by the UE for the following process, i.e., equalization, to assist the bit stream recovery.

\section{Belief Information Based Deep Channel Estimation}
\begin{figure*}[t]
\centering
\includegraphics[width=2\columnwidth]{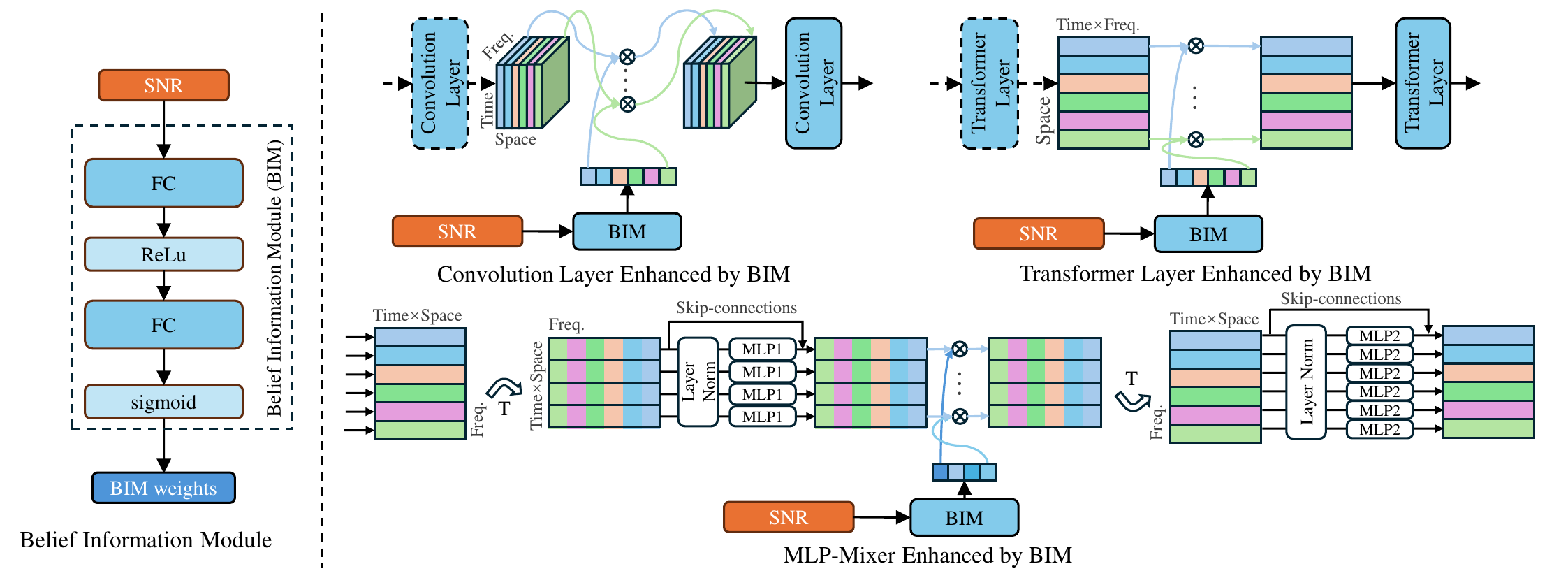}
\caption{The structure of Belief Information Module and structures of BIM enhanced convolution layer,  transformer layer, and MLP-Mixer layer.}
\label{Fig:BIM}
\end{figure*}

Compared with traditional channel estimation methods, DL-based channel approaches can achieve higher channel estimation accuracy and thus reduce pilot overhead. However, to the best of the authors' knowledge, existing DL-based channel estimation methods are primarily tailored for single-antenna channel estimation, despite the deployment of multiple antennas at the UE side in 5G and beyond systems. In this section, we design a belief information based mechanism that fully explores the spatial correlation among multiple antennas. This allows us to extend existing single-antenna channel estimation networks to those suitable for multiple antennas, enhancing overall channel estimation accuracy.

\subsection{Belief Information}
\label{bi}
Owing to its simplicity, most DL-based channel estimation methods utilize the LS method at pilot positions to obtain the initial channel estimate. For a data stream, the LS estimate of the $i$-th receiving antenna is expressed as follows:
\begin{equation}
G^{(i)}_{LS} = \frac{Y^{(i)}_{p}}{X^{(i)}_{p}},
\end{equation}
and the mean square error (MSE) of the LS channel estimate is expressed as:
\begin{equation}
\label{mse}
MSE^{(i)} = \frac{\sigma_z^{2(i)}}{\sigma_x^{2(i)}}=\frac{1}{SNR^{(i)}},
\end{equation}
where $\sigma_x^{2(i)}$ and $\sigma_z^{2(i)}$ denote the $i$-th power of the signals and the the $i$-th power of the noise, respectively. Note that the $MSE^{(i)}$ in Eq.~\eqref{mse} is inversely proportional to the signal-to-noise ratio (SNR) $\sigma_x^{2(i)}/\sigma_z^{2(i)}$. 
The MSE serves as an indicator of the reliability of $G^{(i)}_{LS}$. Specifically, when the equivalent channel $G^{(i)}$ is in good condition, i.e., high SNR, the small MSE suggests that  $G^{(i)}_{LS}$ closely approximates its true value with high probability. Conversely, a low SNR leads a high MSE, increasing the uncertainty of $G^{(i)}_{LS}$. Here, we define the SNR of the $i$-th antenna as its belief information.

Considering the existing DL-based channel estimation methods for a single antenna, the straightforward approach is to adjust their input (i.e., the LS estimate $G_{LS}$) and output (i.e., the equivalent channel $G$) to suit the multiple-antenna scenario, learning their weights with a data-driven manner. However, without integrating communication knowledge, i.e., belief information, these adapted channel estimation networks struggle to fully exploit the spatial relationships among multiple antennas. 

Belief information is extensively utilized to enhance the performance of conventional communication systems, such as employing MRC to improve the performance of the receiver \cite{prakash2011effects}, and using the belief propagation (BP) algorithm for MIMO detection \cite{yoon2013low}. Drawing inspiration from the MRC algorithm, which manually allocates different weights to different antennas based on the belief information to improve the quality of receiving signals, we introduce the belief information into the channel estimation domain. Specifically, the accuracy of the LS estimate elements with lower belief information may be improved by neighboring elements with higher belief information, leveraging the spatial correlation within the same data stream. Nonetheless, it is non-trivial to manually design enhancements for the DL-based channel estimation. In the subsequent subsection, we develop a belief information module and incorporate it into different types of elements in the channel estimation networks.

\subsection{Belief Information Module}
\label{bim}
As shown in the left part of Fig.~\ref{Fig:BIM}, the belief information module (BIM) is expressed as follows:
\begin{equation}
    S = \sigma(W_2\delta(W_1I+b_1)+b_2 \in \mathbb{R}^{c}.
\end{equation}
where $I=[\mu_1, \mu_2, \cdots, \mu_{N_r}] \in \mathbb{R}^{N_r}$ serves as the input of the BIM module, and $S \in \mathbb{R}^{c}$ is the output with $c$ dimensions. Each $\mu_i$ denotes the SNR of the $i$-th receiving antenna, which can be calculated using Eq.~\eqref{mse}. The BIM comprises two fully connected (FC) layers, each equiped with activation functions. $W_1$ and $b_1$ are the weights and the bias of the first FC layer, while $W_2$ and $b_2$ are the weights and the bias of the second FC layer. The activation functions $\sigma$ and $\delta$ represent the ReLU and sigmoid functions, respectively. 

It is worth noting that we do not adopt complex network architectures for the BIM. This simple and efficient structure has been demonstrated effective in our previous research \cite{xu2021wireless}, where we design an attention feature (AF) module. The AF module dynamically recalibrate the features extracted by the convolution layer, considering both the source and channel information. Different from our previous work, based on the analysis of belief information in Section \ref{bi}, we focus solely on the belief information\textemdash specifically, the SNRs of multiple antennas\textemdash as inputs, while maintaining a network structure similar to that used in the AF module.  

As summarized in \cite{sun2023ai}, the channel estimation task can be decomposed into the denoising sub-task and the dimension expansion sub-task. Various backbone architectures could be used for either denoising or dimension expansion, or in a joint manner. Typical backbones, e.g., U-Net, Feature Pyramid Networks (FPN), Uformer, and MLP-mixer, consist of one or several types of elements, e.g., convolution layer, transformer layer, and MLP-mixer layer. The right part of Fig.~\ref{Fig:BIM} illustrates how the BIM is integrated into these layers, specifically adjusting their spatial related features. 

\section{Experimental Results}
\begin{table}[!tb]
\caption{Configuration Parameters}
\label{Table:cp}
\centering
\begin{tabular}{c|c}
\hline
\textbf{Parameters} & \textbf{Values}
\\\hline
Carrier frequency & 2.6 GHz
\\\hline
Sub-carrier spacing & 15 KHz
\\\hline
$N_{t}$ antennas at the BS  & 32
\\\hline
$N_{r}$ antennas at the UE  & 8
\\\hline
$N_{s}$ antennas at the UE  & 1
\\\hline
$N_{c}$ sub-carriers  & 48
\\\hline
$N_{l}$ OFDM symbols  & 14
\\\hline
$N_{cp}$ pilots in frequency domain  & 24
\\\hline
$N_{lp}$ pilots in time domain & 2
\\\hline
Transmission direction & downlink
\\\hline
UE speed & 20 m/s
\\\hline
Modulation & QPSK
\\\hline
Precoding & singular value decomposition (SVD)
\\\hline
\end{tabular}
\end{table}

The following experiments are conducted in the 5G urban macro (UMa) scenario. To demonstrate its compatibility, we select 5G mapping type A configuration type 1 signal-symbol demodulation reference signal (DMRS) as the pilot signals. The DMRSs are spaced by 2 subcarriers and are located in the $2^{nd}$ and the $11^{th}$ OFDM symbols within an OFDM slot. The Detailed configuration parameters are provided in Table \ref{Table:cp}. For network training, we generate 400,000 training samples and 20,000 validation samples by Sionna \cite{sionna}, under a uniform distribution over the EbNo range of [-20, 0] dB. EbNo denotes the ratio of transmit energy per bit to the receiving noise power spectral density. In the evaluation stage, 10,000 test samples are tested at every single EbNo value. Normalized MSE (NMSE) serves as the performance metric.


As detailed in Section \ref{bim}, the BIM is designed to enhance the primary elements of channel estimation networks. In this study, we adapt three channel estimation networks to accommodate the task of multiple-antenna channel estimation and integrate the proposed BIM with them. The first network, ReEsNet \cite{li2019deep}, is a fully convolutional network. The input of ReEsNet is extended from ($N_{cp}$, $N_{lp}$, 2) to ($N_{cp}$, $N_{lp}$, 2$\times N_{r}$) and the output from ($N_{c}$, $N_{l}$, 2) to ($N_{c}$, $N_{l}$, 2$\times N_{r}$). The number of convolution channels is also extended to 96 to boost its channel estimation ability. The second network, Channelformer \cite{luan2023channelformer}, is composed of one transformer layer and convolution layers. Besides the extension of the input shape and output shape, the transformer layer needs two-dimensional data to deal with. Following the method of coupling LS matrices with the transformer layer in \cite{luan2023channelformer}, the frequency dimension (i.e., $N_{cp}$) and the time dimension (i.e., $N_{lp}$) are grouped as one dimension shown in Fig.~\ref{Fig:BIM}(c), and the channels are increased from 2 (representing the real part and the image part) to 2$\times N_{r}$. To modify the last network CMixer \cite{chen2024channel}, we extend its input and output and group the time dimension and space dimension together as shown in Fig.~\ref{Fig:BIM}(d). To simplify terminology, the modified networks retain their original names, and the networks enhanced by BIM are called ``BReEsNet'', ``BChannelformer'' and ``BCMixer''. Adam is adopted to optimize the network during the training stage. The batch size, learning rate, and training epoch are set to 128, 0.001, and 100, respectively.

\begin{figure}[t]
\centering
\includegraphics[width=0.7\columnwidth]{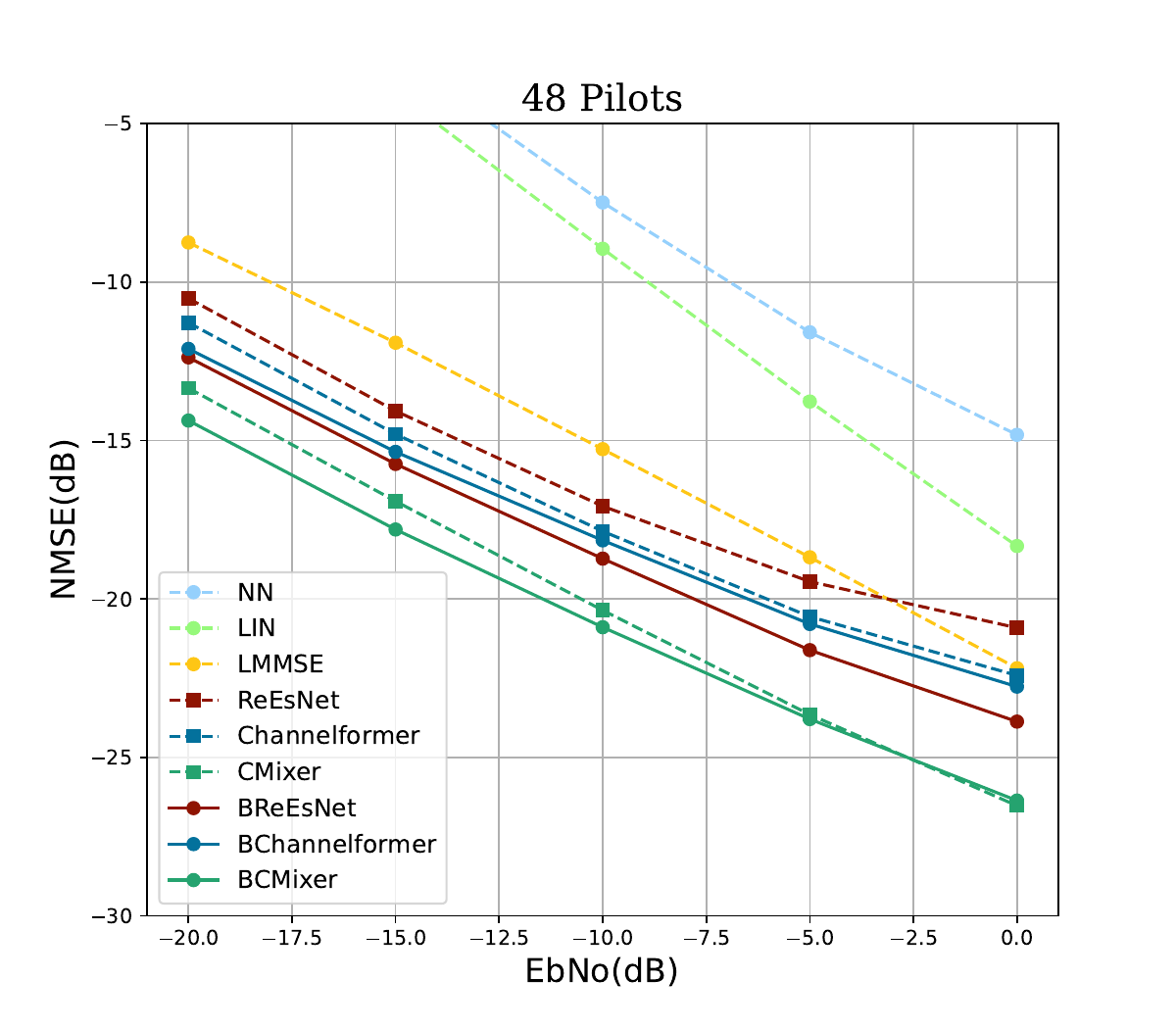}
\caption{Performance of BIM enhanced channel estimation networks and basic channel estimation networks. 
}
\label{Fig:NMSE}
\end{figure}

\begin{table}[!tb] 
\caption{NMSE Performance with different pilot Numbers} 
\label{Table:NMSE_pnum}
\centering
\begin{tabular}{c|c|c|c|c} 
\hline
\multirow{2}*{\textbf{Networks}} & \multicolumn{4}{c}{\textbf{Pilot Number}}\\\cline{2-5}
~&\textbf{48}&\textbf{40}&\textbf{32}&\textbf{24}
\\\hline\hline
ReEsNet & \textbf{-10.52 dB} & -9.98 dB & -9.81 dB & -9.26 dB \\\hline
BReEsNet & -12.38 dB & -11.76 dB & -11.61 dB & \textbf{-11.16 dB} \\\hline\hline
Channelformer  & \textbf{-11.28} dB & -10.63 dB & -10.10 dB & -9.05 dB \\\hline
BChannelformer  & -12.21 dB & -11.48 dB & \textbf{-11.23} dB & -10.55 dB \\\hline\hline
CMixer & \textbf{-13.35} dB & -12.70 dB & -12.11 dB & -11.17 dB \\\hline
BCMixer & -14.35 dB & -13.89 dB & \textbf{-13.26} dB & -12.32 dB \\\hline
\end{tabular}
\end{table}

Fig.~\ref{Fig:NMSE} compares the channel estimation networks enhanced by BIM with the basic channel estimation networks with 48 pilots. Conventional channel estimation methods, i.e., NN, LIN, and LMMSE, serve as references. With the integration of belief information, the performance of BReEsNet, BChannelformer, and BCMixer are 2 dB, 1 dB, and 1 dB better than that of ReEsNet, Channelformer, and CMixer at $\rm EbNo = -20$ dB, respectively. Compared with Channelformer, the performance gain of BChannelformer is reduced with the increase of EbNo. A similar trend is observed when comparing BCmixer with CMixer. Conversely, the inverse tendency is that the performance gain increases when comparing BReEsNet with ReEsNet, potentially due to the use of different backbone networks in these comparisons. Moreover, as illustrated in Table \ref{Table:NMSE_pnum}, we compare the performance of these networks with different pilot numbers at $\rm EbNo = -20$ dB. With the introduction of belief information, the performance of BReEsNet with 24 pilots is even better than that of ReEsNet with 48 pilots. The performance of BChannelformer and the BCMixer with 32 pilots is comparable with that of Channelformer and CMixer with 48 pilots, respectively. These results suggests that the proposed method could reduce $1/3 \sim 1/2$ pilot overhead at low EbNos. 

\begin{figure}[t]
\centering
\includegraphics[width=0.7\columnwidth]{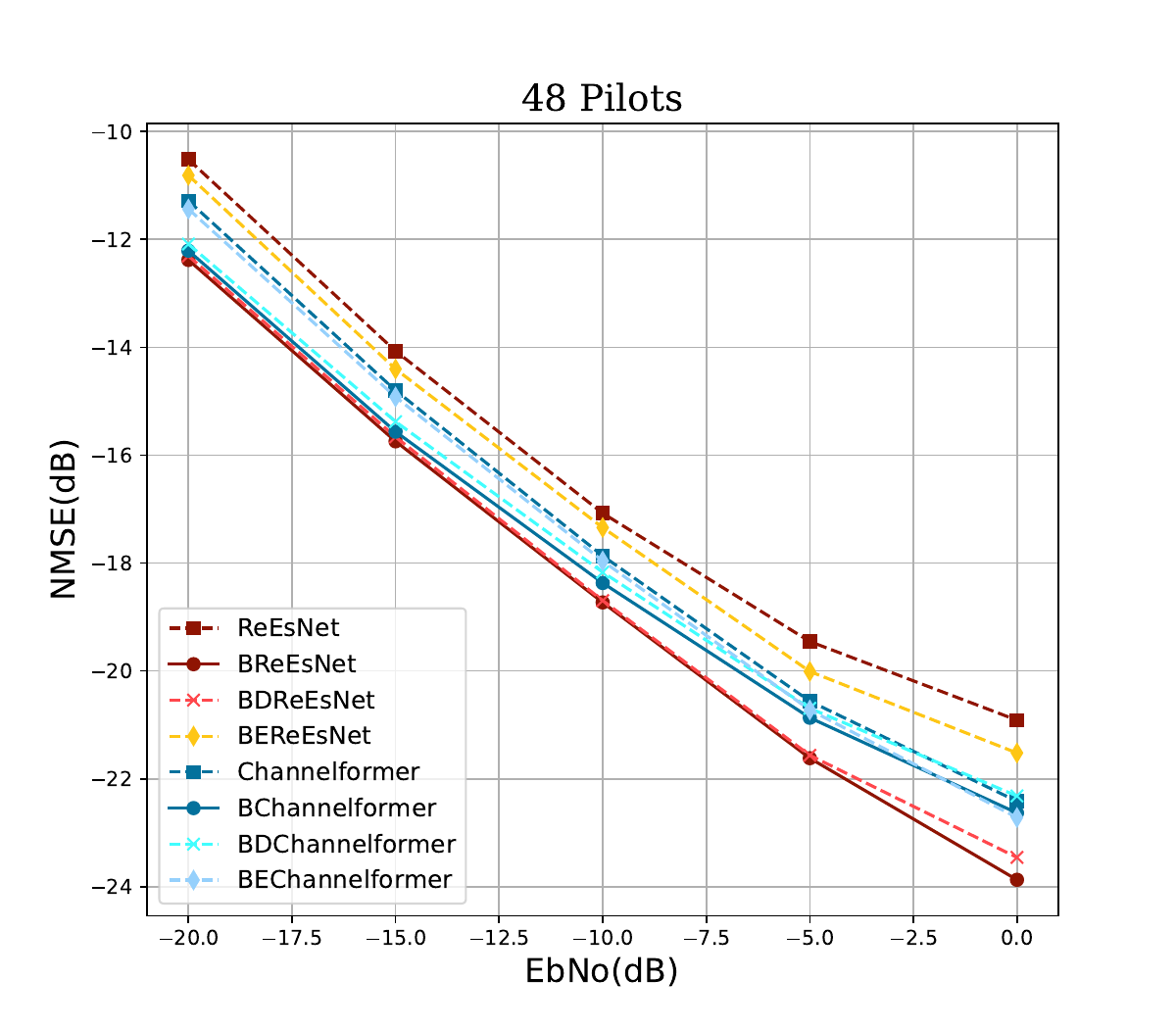}
\caption{Performance of BIM enhanced denoising sub-networks and BIM enhanced expansion sub-networks.}
\label{Fig:Denoise}
\end{figure}

Given that ReEsNet and Channelformer are categorized as ``denoising first'' channel estimation networks, we separately introduce the BIM into their denoising sub-networks (BDReEsNet and BDChannelformer) and expansion sub-networks (BEReEsNet and BEChannelformer). As illustrated in Fig.~\ref{Fig:Denoise}, the performance of BDReEsNet and BDChannelformer closely match that of BReEsNet and BChannelformer, respectively. The reason is that the belief information is directly related to the low-resolution $H_{LS}$ and fully explored by the denoising network. Interestingly, even without introducing belief information into the denoising network in BEReEsNet and BEChannelformer, their performance of still surpasses that of ReEsNet and Channelformer, respectively. This improvement indicates that belief information also plays a beneficial role during the expansion phase. Consequently, we can reasonably infer that the BIM would also enhance ``expansion first'' channel estimation networks.

\begin{table}[!tb]
\caption{Storage overhead and computational Complexity}
\label{Table:socc}
\centering
\begin{tabular}{c|c|c|c}
\hline
\textbf{Methods} & \textbf{NMSE} & \textbf{Params} & \textbf{FLOPs}
\\\hline\hline
ReEsNet (48 pilots) & -10.52 dB & 1,890 K & 198.76 M
\\\hline
BReEsNet (48 pilots) & -12.38 dB & 1,911 K & 198.93 M
\\\hline
BReEsNet (24 pilots) & -11.16 dB & 1,911 K & 108.85 M
\\\hline\hline
Channelformer (48 pilots) & -11.28 dB & 65 K & 53.06 M
\\\hline
BChannelformer (48 pilots) & -12.21 dB & 69 K & 53.16 M
\\\hline
BChannelformer (32 pilots) & -11.23 dB & 55 K & 37.65 M
\\\hline\hline
CMixer (48 pilots) & -13.35 dB & 292 K & 25.27 M
\\\hline
BCMixer (48 pilots) & -14.35 dB  & 307 K & 25.92 M
\\\hline
BCMixer (32 pilots) & -13.26 dB  & 306 K & 25.70 M
\\\hline
\end{tabular}
\end{table}

Table \ref{Table:socc} lists the storage overhead and the computational complexity of the proposed networks and baseline networks. When using the same number of pilots, e.g., 48 pilots, the storage overhead and the computational complexity of the proposed networks are only $0.3\% \sim 6\%$ and $0.1\% \sim 2.6\%$ more than those of the baseline networks, offering enhanced estimation performance. At a similar performance level at $\rm EbNo = -20$ dB, the storage overhead of the proposed networks is comparable to that of the baseline networks, while their computational complexity is reduced by $29.04\% \sim 45.24\%$ (except BCMixer with 32 pilots) compare to the baselines, offering the benefit of reduced pilot overhead.

\section{Conclusion}
We have proposed the belief information based channel estimation mechanism that leverages the belief information extracted from multiple antennas to efficiently explore the spatial relationships among them. In the experiments, we have demonstrated that the proposed BIM is a plug-and-play module, which is easily integrated into existing channel estimation networks and improves channel estimation accuracy. Simulation results indicate that the proposed method could achieve the performance gain or pilot overhead reduction, particularly in environments with low EbNos. In future work, promising directions involves extending the application of the proposed BIM to the task of DL-based joint channel estimation and detection, as well as to DL-based receiver designs.

\bibliographystyle{IEEEtran}
\bibliography{ref/ref.bib}

\end{document}